\documentclass[prl,showpacs,draft,twocolumn,superscriptaddress]{revtex4}

\usepackage[final]{graphicx}

\begin{document}

\title{Asymptotically exact dispersion relations for collective modes
  in a confined charged Fermi liquid}

\author{I.~V.~Tokatly}
\thanks{On leave from Moscow Institute of Electronic Technology,
         Zelenograd, 103498, Russia}

\email{ilya.tokatly@physik.uni-erlangen.de}

\affiliation{Lerhrstuhl f\"ur Theoretische Festk\"orperphysik,
  Universit\"at Erlangen-N\"urnberg, Staudtstrasse 7/B2, 91058
  Erlangen, Germany}

\author{O.~Pankratov}

\affiliation{Lerhrstuhl f\"ur Theoretische Festk\"orperphysik,
  Universit\"at Erlangen-N\"urnberg, Staudtstrasse 7/B2, 91058
  Erlangen, Germany} 
\date{\today}

\begin{abstract} 
Using general local conservations laws we derive dispersion relations
for edge modes in a slab of electron liquid confined by a symmetric
potential. The dispersion relations are exact up to $\lambda^{2}
q^{2}$, where $q$ is a wave vector and $\lambda$ is an effective
screening length. For a harmonic external potential the dispersion
relations are expressed in terms of the {\em exact} static pressure
and dynamic shear modulus of a homogeneous liquid with the density
taken at the slab core. We also derive a simple expression for the frequency
shift of the dipole (Kohn) modes in nearly parabolic quantum dots in a
magnetic field.
\end{abstract}


\maketitle
The examples of importance of a collective motion of particles range from
such old problems as energy losses by fast particles in solids to
modern studies of a collective response in semiconductor nanostructures
and interaction of clusters and molecules with intense laser
radiation. However, only a few {\em exact} features of collective dynamics
in spatially inhomogeneous many-body systems have been found up to
now. The most known result is the Generalized Kohn Theorem (GKT)
\cite{GKT} or the Harmonic Potential Theorem in a more general
formulation \cite{HPT}. This theorem has been extensively used for
construction of approximations for exchange-correlation (xc) potential
in Time Dependent Density-Functional Theory (TDDFT) (see e.g.
Ref.~\onlinecite{TDDFT} and references therein).

In this paper we present another {\it asymptotically exact} result
which, in some sense, can be viewed as an extension of GKT.  We derive
the exact dispersion relations for the dipole and
monopole plasma modes in a slab of a charged Fermi liquid confined by
a symmetric external potential. This
result opens a possibility for a direct experimental determination of
the dynamic shear modulus of an electron liquid. It can also help to
control the accuracy of approximations used to
describe dynamics of inhomogeneous systems. It has also direct
implications for the theory of surface plasmons in simple metals. In
addition we derive a few asymptotically exact relations for dipole
modes in quantum dots (QD) and wires in a homogeneous magnetic field.  

The collective motion of an arbitrary interacting system
obeys the exact local conservation laws:
\begin{eqnarray}
&& \partial_{t}n + \partial_{\mu}n v_{\mu} =0,
\label{1}\\
&&mnD_{t}v_{\mu} + en({\bf v}\times{\bf H})_{\mu} 
+  \partial_{\nu}P_{\mu \nu} + n\partial_{\mu}U = 0,
\label{2}
\end{eqnarray}
where $D_{t}=\partial_{t} + v_{\nu}\partial_{\nu}$, $n$ is the density
of particles, ${\bf v}={\bf j}/n$ is the velocity,
$U=U_{\text{ext}}+U_{\text{H}}$ is a sum of the external and Hartree
potentials, ${\bf H}$ is the external static magnetic field and
$P_{\mu \nu}$ is the exact (generally unknown) stress tensor.

Introducing the displacement
vector ${\bf u}$ as $v_{\mu}=\partial_{t}u_{\mu}$, we derive from
Eq.~(\ref{2}) the linearized equation of motion  
\begin{eqnarray}
\nonumber
mn_{0}\partial_{t}^{2}u_{\mu} 
&+& en_{0}(\partial_{t}{\bf u}\times{\bf H})_{\mu}\\
&+& \partial_{\nu}\delta P_{\mu \nu} 
+  n_{0}\partial_{\mu}\delta U + \delta n\partial_{\mu}U_{0} = 0,
\label{7}
\end{eqnarray}
where $\delta n = -\partial_{\mu}n_{0}u_{\mu}$, $\delta P_{\mu \nu}$
and $\delta U$ are deviations 
of the density, the stress tensor and the potential from their equilibrium
values $n_{0}$, $ P^{\text{eq}}_{\mu\nu}$ and $U_{0}$
respectively. Eigenfrequencies of Eq.~(\ref{7}) define the
exact excitation energies of a system.

Let us first consider the dipole modes of a finite system confined in D
dimensions by the potential $U_{\text{ext}}$ in the presence of a homogeneous
magnetic field ${\bf H}$. Integration of Eq.~(\ref{7}) leeds to the
following exact equation for the dipole moment per particle ${\bf
d}=\frac{1}{N}\int {\bf u}n_{0}d{\bf x}$
\begin{equation}
-m\omega^{2}d_{\mu} = i\omega e({\bf d}\times{\bf H})_{\mu} - \frac{1}{N}\int
n_{0}u_{\nu}\partial_{\mu}\partial_{\nu}U_{\text{ext}}d{\bf r}.
\label{1a}
\end{equation}
This equation, commonly called the dynamical force sum rule
\cite{Sorbello}, shows the exact decoupling of the
center-of-mass and relative motions for an external potential of the
form $U_{\text{ext}}= \frac{1}{2}m\Omega^{2}_{\mu\nu}x_{\mu}x_{\nu}$
\cite{GKT,HPT}. In this case the dipole (Kohn) modes correspond to the
rigid motion with the displacement ${\bf u}= {\bf d}=const$ being the
solution to the Newton equation \cite{HPT}
\begin{equation}
-m(\omega^{2}\delta_{\mu\nu} - \Omega^{2}_{\mu\nu})d_{\nu} = 
i\omega e({\bf d}\times{\bf H})_{\mu}. 
\label{2a}
\end{equation}
In practice one often deals with the quasiparabolic QD, where the
deviation $\Delta U_{\text{ext}}$ of the confining potential from the
parabolic one is small (see
Ref.~\onlinecite{QD} and references therein). The perturbative treatment of
Eq.~(\ref{1a}) shows that to the first order in $\Delta
U_{\text{ext}}$ the dipole moment still satisfies Eq.~(\ref{2a}) but with
the renormalized dynamical matrix $\overline{\Omega}^{2}_{\mu\nu} =
\Omega^{2}_{\mu\nu} + \Delta\Omega^{2}_{\mu\nu}$, where
\begin{equation}
 \Delta\Omega^{2}_{\mu\nu} = \frac{1}{N}\int
n_{0}\frac{\partial^{2}}{\partial x_{\mu}\partial
  x_{\mu}}\Delta U_{\text{ext}} d{\bf r}
\label{3a}
\end{equation}
The corrections to the Kohn-mode frequency has been extensively
studied experimentally  
and theoretically (using approximate many-body methods)
\cite{QD}. The importance of Eq.~(\ref{3a}) in this context is that it
{\em exactly} accounts for the many body effects at the linear 
in $\Delta U_{\text{ext}}$ level. Therefore it can be helpful for
studying correlation phenomena such as Wigner
crystallization in a few-electron QD
(see e.g. Ref.~\onlinecite{WC-FL}) by means of far-infrared
photoabsorption. The transition to the correlated state, which can be 
tuned by a magnetic field, changes the
density distribution $n_{0}$ and consequently affects the frequency shift
according to Eq.~(\ref{3a}).

Another useful relation for a dipole mode can be obtained for a system
of $N$ charged particles confined by a D-dimensional potential
(D$=3,2,1$) with a spherical (dot), cylindrical (wire) or a slab
geometry (we consider the case ${\bf H}=0$). We assume that
$n_{0}(r)$ acquires a constant value $\bar{n}$ inside a system and
vanishes at infinity. The length $\lambda_{0}$ of the density
variation at the boundary is assumed to be smaller than the system
size $a$, which is defined in such a way that the rectangular density
distribution $\bar{n}\theta(a-r)$ provides the correct number of
particles $N$. Noting that the density fluctuation $\delta n({\bf r})$
for the dipole mode is localized at the length scale
$\lambda_{\delta}\ll a$ near the system edges, we obtain from
Eq.~(\ref{1a}) the following result
\begin{equation}
 \omega^{2}-\omega_{p}^{2}/{D} 
= \frac{1}{ma^{D}}\int_{0}^{\infty}
\widetilde{\delta n}(r)
\frac{\partial \Delta U_{\text{ext}}}{\partial r} 
r^{D-1}dr,
\label{4a}
\end{equation}
where $\widetilde{\delta n}(r)=\delta n(r)/\int_{0}^{\infty}\delta
n(r)dr$ is the normalized radial part of the density fluctuation and
$\Delta U_{\text{ext}} = U_{\text{ext}} - U_{\text{har}}$, where the
harmonic potential $U_{\text{har}} = m\omega_{p}^{2}r^{2}/2D$
corresponds to the same equilibrium density in the internal region
($\omega_{p}^{2} = 4\pi e^{2}\bar{n}/m$). It is noteworthy that
$\Delta U_{\text{ext}}$ can be arbitrary lage in the edge region as
the smallness of the right hand side in Eq.~(\ref{4a}) is controlled
by $\lambda_{\delta}/a$.

Equation (\ref{4a}) describes the the frequency shift relative to the
classical value $\omega_{p}/\sqrt{D}$. The shift is either positive or
negative depending on whether $U_{\text{ext}}$ is ``harder'' or
``softer'' than $U_{\text{har}}$. It obviously vanishes for $\Delta
U_{\text{ext}}=0$ in accordance with GKT. Since $\delta n(z)$ has an
imaginary part, the collective mode decays due to Landau damping.

For a wire or a slab geometry Eq.~(\ref{4a}) gives the frequency at
zero wave vector ${\bf q}$. Now we consider a general case ${\bf q}\ne
0$, which requires a local treatment of the basic Eq.~(\ref{7}).

Consider a slab with a given sheet density $N_{s}$
confined in $z$-direction by a symmetric potential
$U_{\text{ext}}(z)$. The equilibrium stress tensor can be written as $
P^{\text{eq}}_{\mu\nu} = P_{0}(z)\delta_{\mu\nu} +
\pi_{\mu\nu}^{\text{eq}}(z)$, where $P_{0}=\frac{1}{3}\text{Tr}
P^{\text{eq}}_{\mu \nu}$ is the exact pressure and $\pi_{\mu
\nu}^{\text{eq}}$ is a traceless tensor which describes quantum
effects. The symmetry requires that $P^{\text{eq}}_{zj}=0$ for $j=x,y$
and $P^{\text{eq}}_{zz} = P_{\perp} = P_{0} + \pi_{0}$,
$P^{\text{eq}}_{xx}=P^{\text{eq}}_{yy} = P_{\parallel} = P_{0} -
\pi_{0}/2$.

The equilibrium density $n_{0}(z)$ satisfies the balance
equation (the static version of Eq.~(\ref{2}))
\begin{equation}
\partial_{z}P^{\text{eq}}_{zz} + 
n_{0}\partial_{z}U_{0} = 0,
\label{4}
\end{equation}
where $U_{0} = U_{\text{ext}} + U_{\text H}$ and $U_{\text H}$ is related to
$n_{0}$ by the Poisson equation
\begin{equation}
\partial_{z}^{2}U_{\text H} = -4\pi e^{2} n_{0}.
\label{5}
\end{equation}
Combining the relation $\delta n = -\partial_{\mu}n_{0}u_{\mu}$ and
Eq.~(\ref{7}) and considering the plane wave solution $e^{-i(\omega t
- {\bf q}{\bf r}_{\parallel})}$ we arrive at equation for the
density fluctuation $\delta n_{{\bf q},\omega}(z)$ 
(indexes ${\bf q},\omega$ are suppressed below)
 \begin{eqnarray}\nonumber
   m\omega^{2} \delta n  
  + \partial_{z}^{2}\delta P_{zz} 
 &-& q_{k}q_{j}\delta P_{kj} 
  + 2q_{j}\partial_{z}\delta P_{zj}\\
  + \partial_{z}\delta n \partial_{z} U_{0}
 &+& \partial_{z} n_{0} \partial_{z}\delta U
  - q^{2} n_{0}\delta U =0,
 \label{8}
 \end{eqnarray}
where $\delta U$ satisfies the equation
\begin{equation}
\left(\partial_{z}^{2} - q^{2} \right)
\delta U(z) = - 4\pi e^{2}\delta n(z).
\label{9}
\end{equation}
Up to this point all transformations are exact. From now on we
consider the equilibrium density $n_{0}(z)$ that satisfies the
assumption used in the derivation of Eq.~(\ref{4a}) \cite{note1}. 

For a slab geometry the width $2a$ is defined as follows  
\begin{equation}
\int_{-\infty}^{\infty}n_{0}(z)dz = 2a\bar{n} = N_{s}.
\label{10}
\end{equation}
Let us choose the origin at the slab ``edge'' so that the
center of the slab is at $z=-a$.
Equation (\ref{10}) implies that the density $n_{0}(z)$ at $z > - a$
can be represented in the form $n_{0}(z) = \bar{n}\theta(-z) + \Delta
n_{0}(z)$, where $\int_{-a}^{\infty}\Delta n_{0}(z)dz = 0$. The
function $\Delta n_{0}(z)$ is localized at the length scale $\lambda_{0}$.

Our assumption allows to determine the asymptotic form of the stress
tensor in the slab 
interior where $n_{0}(z) \approx \bar{n}$ and the gradients of the
density vanish. Here $\pi_{\mu \nu}^{\text{eq}}\approx 0$ and the
equilibrium stress tensor reduces to the pressure
$P^{\text{eq}}_{\mu\nu}\approx \bar{P_{0}}\delta_{\mu\nu}$ (all
quantities marked with a bar refer to the slab interior). In the same region
the local nonequilibrium correction to the stress tensor $\delta
P_{\mu\nu}$ is defined by the asymptotic expression
\begin{equation}
\delta P_{\mu \nu} = - \bar{K}_{\omega}\delta_{\mu \nu}\partial {\bf u} \\
 - \bar{\mu}_{\omega}\left(\partial_{\mu}u_{\nu} + \partial_{\nu}u_{\mu} -
\frac{2}{3}\delta_{\mu \nu}\partial {\bf u} \right)
\label{3}
\end{equation}
Within the limits of Landau Fermi liquid theory the bulk
($\bar{K}_{\omega}$) and the shear ($\bar{\mu}_{\omega}$) moduli are
real and can be expressed in terms of Landau parameters
\cite{Vignale2,TP}. In general $\bar{K}_{\omega}$ and
$\bar{\mu}_{\omega}$ have imaginary parts which are related to
internal viscosities \cite{Vignale2}.

Now we focus on two collective modes at the edges of the
slab. These are the lowest antisymmetric (dipole) and symmetric
(monopole) plasma oscillations. At $q \to 0$ the dipole mode
corresponds to the oscillations across the slab (which is the
Kohn mode in the case of a parabolic confinement), whereas the
monopole mode is a periodic in $q$-direction compression and expansion
of the slab (2D plasmon). At $2qa > 1$ the both modes merge into two
surface plasmons. 

We derive the dispersion relations for these modes with an accuracy
$\lambda^{2}q^{2}$ for arbitrary values of the parameter $aq$, where
$\lambda=\max\{\lambda,\lambda_{\delta}\}$. Let us define an
intermediate length scale $a>l \gtrsim \lambda$ and integrate
Eq.~(\ref{8}) over the region $(-l,\infty)$. The result of the
integration is
 \begin{eqnarray}\nonumber
 \omega^{2} &=& \frac{\omega_{p}^{2}}{2} \left(1 \pm  e^{-2qa}\right)
\left[
 1 - q\langle z\widetilde{\delta n}\rangle -
 \frac{q^{2}}{2}\langle z^{2}\widetilde{\delta n}\rangle 
\right] \\
 &-& \frac{q^{2}}{m}
\left[
 \frac{2\bar{\mu}_{\omega}}{\bar{n}} 
 + \frac{\langle \delta P_{\parallel}\rangle}{\langle\delta n\rangle}
 - D
\right]
 +  O(\lambda^{3}q^{3}),
 \label{12}  
 \end{eqnarray}
where $\langle ...\rangle = \int_{-l}^{\infty}...dz$ and
$\widetilde{\delta n(z)} =\delta n(z)/\langle\delta n\rangle$. The
upper(lower) sign corresponds to the dipole(monopole) mode
and the function $D$ is defined as follows
\begin{eqnarray}\nonumber 
 D &=& 2\pi e^{2}
\big\{ 
 \bar{n}\langle\theta(-z) z^{2}\widetilde{\delta n}\rangle\\
 &+& \langle\langle
 |z-z'|\Delta n_{0}(z)\widetilde{\delta n(z')}
 \rangle\rangle 
 \mp e^{-2qa}\langle z\Delta n_{0}\rangle
\big\}
\label{13}  
\end{eqnarray}
Since both $\Delta n_{0}$ and $\delta n$ vanish at $- z \sim \lambda
\lesssim l$, all integrals in Eqs.~(\ref{12}) and (\ref{13}) are
independent of the lower limit. Equation (\ref{12}) is, in fact, 
an implicit dispersion relation. 

To simplify integrals $\langle z\delta n\rangle$ and $\langle
z^{2}\delta n\rangle$ in Eqs.~(\ref{12}), (\ref{13}), we multiply
Eq.~(\ref{8}) by $z$ and $z^{2}$ respectively and integrate over $z$
from $-l$ to $\infty$. After calculations we arrive at the following
equations
 \begin{eqnarray} \nonumber 
 m\big(\omega^{2} &-& \omega_{p}^{2} \big)
   \langle z\widetilde{\delta n}\rangle 
 = \langle
  \widetilde{\delta n}\partial_{z}\Delta U_{\text{ext}}\rangle\\ 
 &-& 4\pi e^{2}q\left(1 \mp  e^{-2qa}\right)
   \langle z\Delta n_{0}\rangle 
 + O(\lambda^{2}q^{2}), 
\label{14} \\ \nonumber
 m\big(\omega^{2}/2 &-& \omega_{p}^{2} \big)
   \langle z^{2}\widetilde{\delta n}\rangle =
  - \langle \delta P_{\perp}\rangle/\langle\delta n\rangle\\
 &+& \langle 
z\widetilde{\delta n}\partial_{z}\Delta U_{\text{ext}}\rangle 
  - D + O(\lambda q),
\label{15}  
\end{eqnarray}
Using Eqs.~(\ref{12})-(\ref{15}) and introducing the notation 
$$f_{\pm} =1 \pm e^{-2qa},$$
we obtain the dispersion relations in the final form 
\begin{equation}
\omega_{\mp}^{2} = \frac{\omega_{p}^{2}}{2}f_{\pm}
 + \frac{q}{m}\frac{f_{\pm}}{f_{\mp}}
\langle\widetilde{\delta n}\partial_{z}\Delta U_{\text{ext}}\rangle
 + \frac{q^{2}}{m}(A_{\mp} + B_{\mp}) 
\label{16}
\end{equation} 
with the following coefficients in the second-order term
\begin{eqnarray}\label{17}
A_{\mp} 
&=& \Big[\frac{\sqrt{2}f_{\pm}}{\omega_{p}f_{\mp}}
\langle
\widetilde{\delta n}\partial_{z}\Delta U_{\text{ext}}
\rangle\Big]^{2}
- \langle z\widetilde{\delta n}\partial_{z}
   \Delta U_{\text{ext}}\rangle,\\ \nonumber
  B_{\mp} &=& 2\bar{\mu}_{\omega}/\bar{n} 
 + (\langle \delta P_{\parallel} 
 + \delta P_{\perp}\rangle)/\langle\delta n\rangle\\
 &+& 2\pi e^{2}\left(
2\langle z\Delta n_{0}\rangle f_{\pm}
 - \langle z^{2}\widetilde{\delta n}\rangle f_{\mp} 
\right).
\label{18}
\end{eqnarray} 
Let us analyze Eqs.~(\ref{16})-(\ref{18}) in more detail.

If $U_{\text{ext}}$ differs from the harmonic potential the main
contribution to the dispersion comes from the first two terms in the
right hand side of Eq.~(\ref{16}). In the limit $2qa \ll 1$ the lower
mode becomes a 2D plasmon with the universal dispersion 
$\omega_{+}^{2} = 2\pi e^{2}N_{s}q/m$. On
the contrary, the frequency of the higher (dipole) mode
\begin{equation}
\omega_{-}^{2} - \omega_{p}^{2}= -qa\omega_{p}^{2} + 
\langle\widetilde{\delta n}\partial_{z}\Delta U_{\text{ext}}\rangle/am
\label{19}
\end{equation} 
contains a non-universal shift which is proportional to the inverse
width of the slab. At $q=0$ Eq.~(\ref{19}) recovers the result of
Eq.~(\ref{4a}) for 
D=1. However a close inspection of
Eqs.~(\ref{12})-(\ref{15}) shows that, for a dipole mode, they are
consistent only for $q >\lambda/a^{2}$, which seemingly makes the
limit $q\to 0$ inaccessible. Yet in the region $1> qa >
\lambda/a$, where Eqs.~(\ref{12})-(\ref{15}) are applicable, the
dispersion Eq.~(\ref{19}) is a linear function of $q$ which is shifted
by a constant as given in Eq.~(\ref{4a}). Thus, despite
Eqs.~(\ref{12})-(\ref{15}) are ill defined at $q\to 0$, the resulting
dispersion relations, Eq.~(\ref{16}), are correct up to $q= 0$.

In the opposite limit $2qa \gg 1$ ($f_{\pm}\to 1$) both modes merge
into antisymmetric and symmetric combinations of surface
plasmons at two slab surfaces with the same frequency
$\omega_{\mp}=\omega_{S}$
\begin{equation}
\omega_{S}^{2} = \omega_{p}^{2}/2 + q \langle\widetilde{\delta
n}\partial_{z}\Delta U_{\text{ext}}\rangle/m.
\label{20}
\end{equation}
In a particular case of the jellium edge $\partial_{z}\Delta
U_{\text{ext}} = -m\omega_{p}^{2}z\theta(z)$ and we recover the result
\begin{equation}
\omega_{s}^{2} = \omega_{p}^{2}/2 - q\omega_{p} \int_{0}^{\infty}z
 \widetilde{\delta n}dz,
\label{21}
\end{equation}
obtained in Ref.~\onlinecite{Liebsch} using the dynamical force sum rule. 

The intriguing feature of 
Eqs.~(\ref{16})-(\ref{18}) is a very special role of the harmonic
potential. Both the linear term in Eq.~(\ref{16}) and one of the second order
coefficients $A_{\mp}$, Eq.~(\ref{17}), vanish if $\Delta
U_{\text{ext}}(z)=0$ \cite{Schaich}. This is obviously related to GKT
\cite{GKT}. 
The existence of the exact GKT mode with $\omega = \omega_{p}$ at $q=0$
can be easily observed via direct substitution of the rigid-shift solution:
$u_{\mu}({\bf r},t)=\delta_{\mu z}u_{z}(t)$, and
\begin{equation}
\delta n({\bf r},t) = - u_{z}\partial_{z}n_{0}, \quad
\delta P_{\mu\nu}({\bf r},t) = - u_{z}\partial_{z}P^{\text{eq}}_{\mu\nu}
\label{22}
\end{equation} 
into Eqs.~(\ref{7}) and (\ref{9}) at $q=0$ with
$U_{\text{ext}}=U_{\text{har}}$. Interestingly,
the solution  Eq.~(\ref{22}) remains
asymptotically valid at finite $q$ for both modes under
consideration. Substituting Eq.~(\ref{22}) with
$u_{z}= u_{z}(t)e^{i{\bf q}{\bf r}_{\parallel}}$ into Eqs.~(\ref{7})
and taking into account the Poisson equation for $\delta U$ we obtain
the equation
\begin{equation}
\partial_{t}^{2}u_{z}(t) + 2\pi e^{2}\bar{n} \left(1 \pm
e^{-2qa}\right)u_{z}(t) + O(qz)=0,
\label{23}
\end{equation}
which shows the validity of the rigid-motion solution in the
region $z \ll 1/q$ near the edges of the slab. This observation allows
to simplify the remaining coefficient $B_{\mp}$ [Eq.~(\ref{18})] in
Eq.~(\ref{16}). Using Eq.~(\ref{22}) we
find that the integrals of the stress tensor fluctuations (the second
term in Eq.~(\ref{18})) are reduced to the equilibrium pressure per
particle in the internal region
\begin{equation}
\langle \delta P_{\perp}\rangle /\langle\delta n\rangle =
\langle \delta P_{\parallel}\rangle /\langle\delta n\rangle =
\bar{P}_{0}/\bar{n},
\label{24}
\end{equation}
whereas the integral $\langle z^{2}\widetilde{\delta n}\rangle$, which
enters the last term in Eq.~(\ref{18}), is transformed to the form
\begin{equation}
2\pi e^{2}\bar{n}\langle z^{2}\widetilde{\delta n}\rangle =
4\pi e^{2} \langle z\Delta n_{0}\rangle.
\label{25}
\end{equation}
The right hand side of Eq.~(\ref{25}) is, in fact, the potential of a
double layer which is formed by the density distribution $\Delta
n_{0}(z)$. On the other hand, the integration of the
balance equation Eq.~(\ref{4}) leads to the following
exact relation of the double layer potential to the pressure per
particle inside the slab
\begin{equation}
4\pi e^{2} \langle z\Delta n_{0}\rangle = \bar{P}_{0}/\bar{n}.
\label{26}
\end{equation}

Substituting Eqs.~(\ref{24})-(\ref{26}) into Eqs.~(\ref{18}) and
(\ref{16}) we obtain {\em explicit} dispersion relations for
collective modes in a parabolically confined system
\begin{equation}
\omega_{\mp}^{2} = 
\frac{\omega_{p}^{2}}{2} f_{\pm} +
\left(\frac{2\bar{P}_{0}}{m\bar{n}}f_{\pm} +
\frac{2\bar{\mu}_{\omega}}{m\bar{n}} \right)q^{2}.
\label{27} 
\end{equation}
In the surface plasmon regime ($2qa > 1$, $\omega_{\mp}=\omega_{S}$)
this equation simplifies further as
\begin{equation}
\omega_{s}^{2}=
\omega_{p}^{2}/2  +
2q^{2}\left(\bar{P}_{0} + \bar{\mu}_{\omega}\right)/m\bar{n} =
\omega_{p}^{2}/2 + v_{S}^{2}q^{2}.
\label{28} 
\end{equation}
This equation relates the dispersion of plasma modes to the exact
pressure $\bar{P}_{0}$ and shear modulus $\bar{\mu}_{\omega}$ of a
homogeneous electron liquid at a given temperature $T$ and density
$\bar{n}$. Equation (\ref{27}) shows the absence of Landau damping up
to $O(\lambda^{2}q^{2})$. The attenuation is due to the internal
viscosity (the imaginary part of $\bar{\mu}_{\omega}$) which is
related to the multi-pair excitations. From exact results
Eqs.~(\ref{27}) or (\ref{28}) we can easily obtain the dispersion in any
particular approximation. For example, in Random Phase Approximation
(RPA), which assumes Hartree approximation at the static level, we
have $\bar{\mu}_{\omega}=\bar{P}_{0}$ \cite{TP} and, therefore,
$v_{S}^{2}=4\bar{P}_{0}/m\bar{n}$ (i.e. $\frac{4}{5}v_{F}^{2}$ at
$T=0$). This is times $4/3$ of the corresponding coefficient in the
RPA bulk plasmon dispersion.

As the exact pressure $\bar{P}_{0}(\bar{n})$ is presently available
from Monte-Carlo calculations, Eq.~(\ref{27}) opens a direct access to
the complex dynamic shear modulus $\bar{\mu}_{\omega}$ in the range $0
< \omega < \omega_{p}/\sqrt{2}$. To determine $\bar{\mu}_{\omega}$ one
has to measure the dispersion of plasma modes in a wide parabolic
quantum wells using e.g. the grating technique \cite{grating}. To
access the quadratic part of the dispersion, which is governed by the
shear stress, one needs thinner grating and, possibly, wider wells
than those used in Ref.~\onlinecite{grating}. Such measurements would
allow a direct comparison to recent many-body calculations
\cite{Conti}.

As any exact result, Eq.~(\ref{27}) should help to control the
consistency of approximate methods such as different hydrodynamical
approaches \cite{Bloch,Dobson1,TP} or TDDFT schemes with approximate
xc kernels $f_{\text xc}$ \cite{TDDFT}.

Let us consider hydrodynamics first. Since the dispersion
Eq.~(\ref{27}) is controlled by the shear stress, any hydrodynamics
with a diagonal stress tensor $\delta P_{\mu\nu}\sim \delta_{\mu\nu}$
\cite{Bloch,Dobson1} does not reproduce the correct dispersion
coefficient. In the adiabatic hydrodynamics \cite{Bloch} also the
dispersion of bulk plasmon is wrong. This problem is removed in the
theory of Ref.~\onlinecite{Dobson1}, which is however valid only for
1D motion, where the tensor structure of $\delta P_{\mu\nu}$ is
irrelevant. The generalized hydrodynamics of Ref.~\onlinecite{TP}
gives the exact result since it reproduces the correct structure of
$\delta P_{\mu\nu}$.
 
Similar arguments apply to different approximations for $f_{\text xc}$
in TDDFT.  The kinetic part of $\mu$, which is equal to the pressure
of Kohn-Sham particles, is reproduced correctly. The xc part is,
however, $f_{\text xc}$-dependent. For example, the adiabatic
approximation gives no additional xc contribution to the shear stress
which is completely a non-adiabatic effect.
The consistent result can be obtained using the approximate xc
kernel of Ref.~\onlinecite{Vignale1}. The correct
asymptotic form of the stress tensor Eq.~(\ref{3}) was, in fact, one of the
requirements in the derivation of this approximation. The dispersion
of surface plasmons is an example of physical situation
where this asymptotic requirement leads to an observable effect.

The work of I.T. was supported by the Alexander von Humboldt
Foundation and in part by the Russian Federal Program ``Integration''.


\begin{thebibliography}{99}
\bibitem{GKT} L.~Brey, N.~F.~Johnson, and B.~I.~Halperin,
  Phys. Rev. {\bf B 40}, 10647 (1989). 
\bibitem{HPT} J.~F.~Dobson, Phys. Rev. Lett {\bf B 73}, 2244 (1994). 
\bibitem{TDDFT} K.~Burke and E.K.U.~Gross, in {\it Density
    functionals: Theory and applications} (Springer, Berlin, 1998).
\bibitem{Sorbello} R.~S.~Sorbello, Solid State Commun. {\bf 56}, 821 (1985).
\bibitem{QD} L.~Jacak, P.~Hawrylak, and A.~Wojs, {\it Quantum Dots}
  (Springer, Berlin, 1997); O.~Astafiev, et.al, Phys. Rev. {\bf B
  65}, 085315 (2002); V.~Gudmundsson and R.~R~Gerhardts, Phys. Rev. {\bf B
  43}, 12098 (1991); C.~A.~Ullrich and G.~Vignale, Phys. Rev. {\bf B
  61}, 2729 (2000).  
\bibitem{WC-FL} C.~Yannouleas and U.~Landman, Phys. Rev. Lett. {\bf
    82}, 5325 (1999); R.~Egger et al., Phys. Rev. Lett. {\bf 82}, 3320
    (1999); S.~A.~Mikhailov, Phys. Rev. {\bf B 65}, 115312 (2002).
\bibitem{note1} The upper standing plasma modes in a similar system
  have been studied in O.~Heinonen and W.~Kohn, Phys. Rev. {\bf B 48},
  12240 (1993).  
\bibitem{Vignale1} G.~Vignale, C.~A.~Ullrich and S.~Conti, Phys.
Rev. Lett. {\bf 79}, 4878 (1997).
\bibitem{Vignale2} S.~Conti and G.~Vignale,  Phys. Rev. {\bf B 60},
  7966 (1999). 
\bibitem{TP} I.~V.~Tokatly and O.~Pankratov, Phys. Rev. {\bf B 62},
  2759 (2000).
\bibitem{Liebsch} A.~Liebsch, Phys. Rev. {\bf B 36},
  7378 (1987)
\bibitem{Schaich} The absence of the linear term in the surface
  plasmon regime was noted in W.~L.~Schaich, Surface Science, {\bf
  318}, L1157 (1994).  
\bibitem{grating} P.~R.~Pinsukanjana, et. al., Phys. Rev. {\bf B 46},
  7284 (1992). 
\bibitem{Conti} R.~Nifosi, S.~Conti, and M.~P.~Tosi, Phys. Rev. {\bf B 58},
  12758 (1998). 
\bibitem{Bloch} F.~Bloch, Z. Phys. {\bf 81}, 363 (1933); E.~Zaremba
  and H.~C.~Tso, Phys. Rev. {\bf B 49}, 8147 (1994). 
\bibitem{Dobson1} J.~F.~Dobson and H.~M.~Le, J. Molecular Structure
  (Theochem) {\bf 501-502} 327 (2000); cond-mat/0201267.
\end{thebibliography}
\end{document}